# $Cu_{1-x}Fe_xO$: Hopping Transport And Ferromagnetism

Mohd. Nasir[a], Md. A Ahmed[b], Rakibul Islam[c], Gautham Kumar[d], Sunil Kumar[e], C L Prajapat[f], Sajal Biring[d,†], Somaditya Sen[a,†]

[a] *Department of Physics, Indian Institute of Technology Indore, 452020, India*

[b] *Department of Physics, University of Calcutta, Kolkata, 700009, India*

[c] *University of Lille-Sciences and Technologies, UFR de Physi Villeneuve d'Ascq, 59655 France*

[d] *Electronic Engg., Ming Chi University of Technology, New Taipei City, 8802, Taiwan*

[e] *Metallurgical Engineering and Material Science, Indian Institute of Technology Indore, 452020, India*

[f] *Technical Physics Division, Bhabha Atomic Research Centre, Mumbai, 400085, India*

**ABSTRACT**

Single phase, sol-gel prepared $Cu_{1-x}Fe_xO$ ($0 \leq x \leq 0.125$) are characterized in terms of structural, electronic and magnetic properties. Using dielectric and magnetic studies we investigate the coupling of electron and spin. The electrical conductivities and activation energies are studied with increasing Fe content. Modeling of experimental conductivity data emphasizes on a single hopping mechanism for all samples except x=0.125, which have two activation energies. Hole doping is confirmed from Hall effect [1] and by confirming a majority $Fe^{3+}$ substitution of $Cu^{2+}$ in CuO from XPS studies. Such a substitution results in stabilized ferromagnetism. Fe substitution introduces variation in coercivity as an intrinsic magnetic property in Fe-doped CuO, and not as a secondary impurity phase.

**† Corresponding Author e-mail:** sens@iiti.ac.in (S Sen)

Dilute magnetic semiconductors (DMS), in which ferromagnetic behavior above room temperature is induced in nonmagnetic semiconductors by doping with magnetic transition-metal (TM) ions, are of great interest as potential semiconductor-compatible magnetic components for practical spintronics applications [2–4]. Based on the theoretical calculations [4] done by Dietl *et al*. investigations on room-temperature ferromagnetism (RTFM) in DMS materials increased since 1990. Oxides such as ZnO, $TiO_2$, $SnO_2$ *etc.* have been and are still being explored as host materials for preparing DMSs [5-10]. However, so far a low Curie temperature restricts these materials for practical applications. Optical transparency, electrical conductivity and ferromagnetism with high $T_C$ are essential properties of an attractive DMS. DMS based on TM-doped CuO could be useful for a variety of applications requiring combined magnetic and optical functionality.

Copper (II) oxide, CuO has a monoclinic structure. It is important due to a rich physics of high critical-temperature superconducting materials whose basic units are Cu-O chains or layers [11]. It is also important as a probable multiferroic due to induced ferromagnetism in doped CuO. Pure CuO, is a p-type antiferromagnetic material with a band gap of 1.7 eV and a Neel temperature of 230 K [12]. It is non-toxic and abundant. CuO is also suitable for its probable application in microelectronics as capacitors and memory devices.

Transition metal doping makes radical changes in optical, electrical, and magnetic properties of CuO by altering its electronic structure, observed experimentally in Fe [13], Ni-doped CuO [14] Fe and Ni co-doped CuO [15]. Wide controversies exist in explaining the observations, the reproducibility and the suitability of preparation methods. Only a few studies on conduction of CuO are available [16]. Keeping in mind that small amount of impurities in the specimen or thermal history plays a major role in determining the electrical and magnetic properties, we investigate the conduction mechanism and correlated magnetic properties to understand RTFM in CuO by Fe doping.

Nanocrystalline $Cu_{1-x}Fe_xO$ ($x$ = 0, 0.027, 0.055, 0.097 and 0.125) powders were prepared by standard Pechini sol-gel process. The synthesis and structural properties were already discussed in detail in our previous report [17]. The electronic properties of the constituent ions were investigated using x-ray photoelectron spectroscopy (XPS) system of SPECS with monochromatic Al-$K_\alpha$ x-ray (hv = 1.48 KeV) radiation as the primary radiation source and having an optimum energy resolution of 0.5 eV. Broadband Dielectric Spectroscopy (Solarton Analytical—Ametek) was employed to investigate the electrical properties of $Cu_{1-x}Fe_xO$. Magnetic field and temperature dependent magnetization of the samples were investigated using Quantum Design SQUID VSM (model SVSM-050).

Structural studies using X-ray Diffraction (XRD) was discussed previously [17] and revealed monoclinic crystalline phase of the $Cu_{1-x}Fe_xO$ nanoparticles for 0<x<0.125. Changes in lattice constants were observed from Reitveld refinement showing decrease in a and b, while c shows a rapid decrease upto x=0.027 and thereafter becomes independent of substitution. The average crystallite size decreased with substitution.

The valence states of the constituent elements in the material were examined by X-ray photo electron spectroscopy (XPS). The broad scan XPS survey of the $Cu_{1-x}Fe_xO$ [Figure 1(a)] shows the presence of core-level lines of Cu, O 1$s$ and Fe. In addition, the C 1$s$ is also observed corresponding to contamination from the air on the surface of pellet samples. The high resolution Cu 2$p$ spectrum (Figure 1(b)) shows a Cu 2$p$ doublet (Cu 2$p_{3/2}$ an Cu 2$p_{1/2}$) at the binding energies of 930.85 and 939.73 eV respectively. There is appearance of a shake-up satellite about

8.88 eV above the main peak (Cu $2p_{3/2}$). This confirms $Cu^{2+}$ valence states. The satellite is a characteristic of materials such as; copper halides having open shell $3d^9L$ (L for ligand) configuration [18]. This gives rise to multiplet splitting in the $2p^53d^9$ final sate.

The high resolution Fe $2p$ spectrum (Figure 1(c)) in the $Cu_{0.875}Fe_{0.125}O$ showing a Fe $2p$ doublet (Fe $2p_{1/2}$ and Fe $2p_{3/2}$) is observed at ~ 722.7 eV and ~710-714 eV, respectively. The splitting in Fe $2p$ spectra having the separation of 12.7 eV is due to spin-orbit coupling. The binding energy of $Fe^{2+}$ occurs in the range 709.4 eV to 710.3 eV and that of $Fe^{3+}$ occurs in the range 710.3 eV to 711.4 eV [19,20]. The less intense broad Fe $2p$ peak around 710 eV to 714 eV indicates that the Fe is present as mixed valance of a majority 3+ and a minority 2+ in the CuO lattice. A shake-up satellite contribution is also observed at about 717.59 eV which is in close agreement with available literature, indicating that Fe is in majority of 3+ valence state [21]. The absence of peak at 706-707 eV rules out the presence of metallic clusters. Hence, magnetism should not be attributed to interstitial or externally existing iron clusters.

A previous study on the same samples ruled out the presence of metallic Fe clusters, FeO and $Fe_2O_3$ phases from SXAS and XANES studies confirming $Fe^{3+}$ incorporation in CuO lattice. EXAFS studies revealed that the local neighborhood of $Fe^{3+}$ matches with that of $Cu^{2+}$ in CuO confirming substitution. Any similarity with oxides of Fe and other complex oxides of Cu and Fe was ruled out. EXAFS data analysis revealed reduction in oxygen vacancies with increasing Fe content possibly to due to extra charge of $Fe^{3+}$ than $Cu^{2+}$ ions within the same structure [17].

The frequency dependent real part σ′ of the complex electrical conductivity σ*(f) is connected to the imaginary part $\mathcal{E}''$ of the complex permittivity $\mathcal{E}$*(f) through the relation [22] σ′ = $2\pi f \mathcal{E}_0 \mathcal{E}''$, where, f denotes the experimental frequency of the harmonic voltage applied to the specimen. Figure 2(a and b) show the real part of the electrical ac conductivity (σ′(f)) for all compositions as a function of frequency at different temperatures (153-293K). The σ′(f) increases with increasing frequency above a characteristic onset frequency $f_H$, below which it is non variant but shows a dispersion at higher frequencies. The non-variant region at very low frequencies can be compared to the dc conductivity $\sigma_{dc}$. There is no significant change in conductivity is noticed in a wide frequency range. The dispersion shifts to lower frequencies with decreasing temperatures. With increasing Fe content, and decreasing temperature σ′(f) decreases.

Hopping carriers interacts with the inherent defects or the disorderedness in the material especially in the low frequency regime. The Jonscher's universal dielectric response (UDR) model [23], originates from such interactions, given by,

$$\sigma'(f) = \sigma_{dc} \left[ 1 + \left(\frac{f}{f_H}\right)^n \right] \qquad (1)$$

where, $\sigma_{dc}$ is the dc conductivity, $f_H$ is onset frequency of the hopping process, and n is a frequency exponent parameter in the range $0 \leq n \leq 1$. The frequency dependent real part of the electrical conductivities of the $Cu_{1-x}Fe_xO$ samples have been modeled [Figure 2(a and b)] using Jonscher's UDR model. The temperature dependence of $\sigma_{dc}$ and $f_H$ were extracted from the above model and has been plotted with temperature [Figure 3]. With increasing temperature, $\sigma_{dc}$ increases non-linearly, revealing semiconducting nature of the samples. Hopping conduction occurs through the neighboring sites in the nearest-neighbor-hopping (NNH) conduction model. The hopping range and activation energy does not depend on temperature [24,25] and can be analyzed by the Arrhenius equation [24],

$$\sigma_{dc}(T) = \sigma_0 \exp(-E_a/K_B T) \qquad (2)$$

where, $\sigma_0$ is a constant representing the dc conductivity at T→∞, $K_B$ is the Boltzmann constant, and $E_a$ is the activation energy for hopping conduction. $\sigma_{dc}$ was fitted with the NNH model. The Arrhenius model can be better understood by plotting $\log(\sigma_{dc}[S.cm^{-1}])$ against $1000/T$ $(K^{-1})$ [Figure 3(b)]. A linear nature of this plot ensures Arrhenius behaviour. It was observed that for lower substitution the plots were extremely linear. However, for x≥0.055, some non-linearity was observed. For x=0.125 the plot was extremely non-linear. The non-Arrhenius behavior is rectified in literature by Vogel-Tamman-Fulcher (VTF) model, given by,

$$\sigma_{dc}(T) = \sigma_0 \exp\left[-\frac{E_a}{K_B(T-T_0)}\right] \qquad (3)$$

where, $T_0$ is the Vogel temperature. In VTF model, the migration of ions depend on such defects which arise from structural modification within the system [26].

The VTF equation provides a good fit to the dc conductivity for all samples. For low x, $T_o$→0, but for x>0.055, $T_o$~150K. The Arrhenius model thus fits well for x≤0.27, but the VTF model dominates beyound x≥0.055. This transformation is most probably due to ionic motion in which Fe plays an important role fascilitated by the decreasing grain size of the crystallites. $Fe^{3+}$ ion having a different ionic radius is most likely a source of internal strain, not only creating localised defects but also reducing single crystal domain sizes within the same $Cu_{1-x}Fe_xO$ grain as already reported by SEM studies [1]. $E_a$ and $\sigma_0$ decreases with increasing substitution [Figure 3 (a); (inset)]. The decreasing trend of $\sigma_{dc}$ of these samples was also directly calculated from I-V characteristics and was reported previously [1].

CuO is a p-type material, and $Fe^{3+}$ ions, provides an excess electron to the lattice. However, due to the significantly p-type CuO host this excess electron cannot improve the conductivity. It was reported from Hall measurements that p-type carrier concentration decreased [1] reducing the net conductivity of the material with increasing substitution. Reduction in oxygen vacancies with increasing Fe content was found from EXAFS analysis [17] due to extra charge of $Fe^{3+}$ than $Cu^{2+}$ ions within the same structure. The reducing domain sizes will generate more domain walls while lattice will be more defected with increasing substitution, thereby reducing mobility of charge carriers. Thus with increasing substitution most probably mobility is reduced. Thus the variation in $\sigma_0$ and $E_a$ is most probably due to a combination of reduced carrier concentration as well as mobility. Both Arrhenius and VTF models are single activation energy models. Good fits to the experimental data using these models emphasize on a single conduction mechanism present in all the substituted samples. However, only for x=0.125, from the $\ln\sigma_{dc}$ vs 1000/T plots it seems that a double activation model may also fit the data with activation energies ~0.2 and 0.4eV.

Similar to the $\sigma_{dc}$, the hopping frequency $f_H(T)$ data is fitted with the Arrhenius model [Figure 3(c and d)]: $f_H(T) = f_0 \exp(-E_H/K_B T)$; where, $f_0$ is a constant and $E_H$ is the activation energy of hopping frequency of carriers, as well as the Vogel-Tamman-Fulcher (VTF) model, defined as, $f_H(T)=f_1\exp[-E_H/k_B(T-T_0)]$, where $f_1$ and $T_0$ are fitting parameters. Similar to the $\sigma_{dc}$ data it is noticed that $f_H$ follow the same trend as a function of temperature. At very low substitution the Arrhenius nature prevails but as substitution increases a VTF model dominates. The $E_H$ and $f_o$ decreases with incresed Fe substitition [Inset; Figure 3(d)].

To explain the relationship between $\sigma_{dc}$ and $f_H$, log($\sigma_{dc}$[S.cm$^{-1}$]) vs log($f_H$[Hz]) is plotted which is shown in Figure 2(c). The linear behavior of the plot follows the Barton-Nakajima-Namikawa (BNN) relation. It is noteworthy that upon Fe substitution the conductivity decreases instead of increasing once again emphasizing on the point that proper substitution has happened in the samples.

The field dependent magnetization [M versus H] curves for $Cu_{1-x}Fe_xO$ samples [Figure 4 (a-c)] at different temperatures are ferromagnetic in nature. Saturation magnetization, $M_S$, is not achieved upto 5000 Oe. The remnant magnetization, $M_R$, and coercive field, Hc, increase linearly with substitution [Figure 4(d-e)] indicating stronger ferromagnetic exchange interaction. The varying nature of Hc and $M_R$ along with $M_S$ with substitution is a proof of magnetization being not due to impurity phase of iron oxides or metallic clusters of Fe but is due to proper substitution.

Oxidation states of Fe ions in CuO plays a critical role in magnetic properties of CuO. It has been mentioned that in the $Fe^{3+}$ substituted samples, the p-type nature is retarded while the oxygen deficiency is reduced. This type of change would not have been expected if it was a $Fe^{2+}$ substitution, the p-type nature might not have altered and the oxygen deficiency not reduced. Oxygen vacancies have played an important role in magnetism by actively contributing metal-oxygen-metal double exchange contribution. A reduction in oxygen vacancy thereby makes such exchange integrals stronger. Hence $Fe^{3+}$ substitution opens up a better chance of $Fe^{3+}$-$O^{2-}$-$Cu^{2+}$ superexchange or a $Fe^{3+}$-$O^{2-}$-$Cu^{2+}$-$O^{2-}$-$Fe^{3+}$ double superexchange phenomena. Therefore, apart from contributing to the electrical properties, this $Fe^{3+}$ substitution also modifies the magnetic properties. Note that the replacement of $Cu^{2+}$ by $Fe^{3+}$ ion in CuO enhances the magnetic moment as the total spin of $Fe^{3+}$ (5/2) is more than that of $Cu^{2+}$ (1/2) ion.

The influence of TM doping on magnetic properties of CuO was studied by Wesselinowa, [27] and it was found that the exchange interaction $J_{ij}=J(r_i-r_j)$ depends on the distance between the spins. The smaller the lattice parameters, shorter are the interionic distances and thereby stronger is the exchange interaction. The ionic radii of $Fe^{3+}$ (0.61 A) is lesser than $Cu^{2+}$ (0.73 A). Hence the lattice tends to contract. XRD results revealed that lattice parameters have reduced with increasing Fe substitution. As calculated by Wesselinowa, the Fe-O-Cu superexchange increases thereby increasing the ferromagnetism in the material. In these samples we observe a proper experimental evidence of the same, and link the results to the role of decreasing oxygen deficiency. The increase of ferromagnetism with decreasing temperature is explainable by reducing lattice parameters. The saturation magnetization does not saturate because of the antiferromagnetic component due to Cu-O-Cu interactions. The increasing remnant magnetization and coercive fields are a result of increasing amount of Fe-O-Cu and Fe-O-□ interactions, where, □ are cation vacancies created due to excess $Fe^{3+}$ substitution. The excess charge of $Fe^{3+}$ ion not only attracts oxygen to reduce oxygen vacancies but beyond a limit to maintain charge balance will generate cation vacancies, □. These exchange interactions are sometimes stronger that a normal $Fe^{3+}-O^{2-}-Cu^{2+}$ superexchange or a $Fe^{3+}-O^{2-}-Cu^{2+}-O^{2-}-Fe^{3+}$ double superexchange interaction, thereby enhancing the ferromagnetism.

Monoclinic single phase $Fe^{3+}$ substituted $Cu_{1-x}Fe_xO$ ($x=$ 0, 0.027, 0.055, 0.097, 0.125) has been examined by XPS, electrical and magnetic studies. Impurity phases of Fe metallic cluster, $Fe_2O_3$ and $Fe_3O_4$ have been ruled out from XPS, other structural studies and magnetic measurements. The electrical conductivities and activation energies are found to decrease with increase in Fe content. The experimental data can be modeled using a single hopping mechanism for all samples except x=0.125, which have two activation energies. Weak ferromagnetic behaviour has been observed at room temperature for all the samples. Magnetism increases with decreasing temperature and increased Fe substitution. The increasing remnant magnetization and coercive fields in substituted CuO results from increasing amount of Fe-O-Cu and Fe-O-□ interactions, where, □ are cation vacancies created due to excess $Fe^{3+}$ substitution. These exchange interactions are sometimes stronger that a normal $Fe^{3+}$-$O^{2-}$-$Cu^{2+}$ superexchange or a $Fe^{3+}$-$O^{2-}$-$Cu^{2+}$-$O^{2-}$-$Fe^{3+}$ double superexchange interaction, thereby enhancing the ferromagnetism. This study motivates a future size dependent magnetic study of Fe-doped CuO.

The authors gratefully acknowledge Professor Pradeep Mathur, Director, IIT Indore for his valuable support to the project. M Nasir is thankful to UGC, New Delhi, for providing Maulana Azad fellowship.

List of Figures

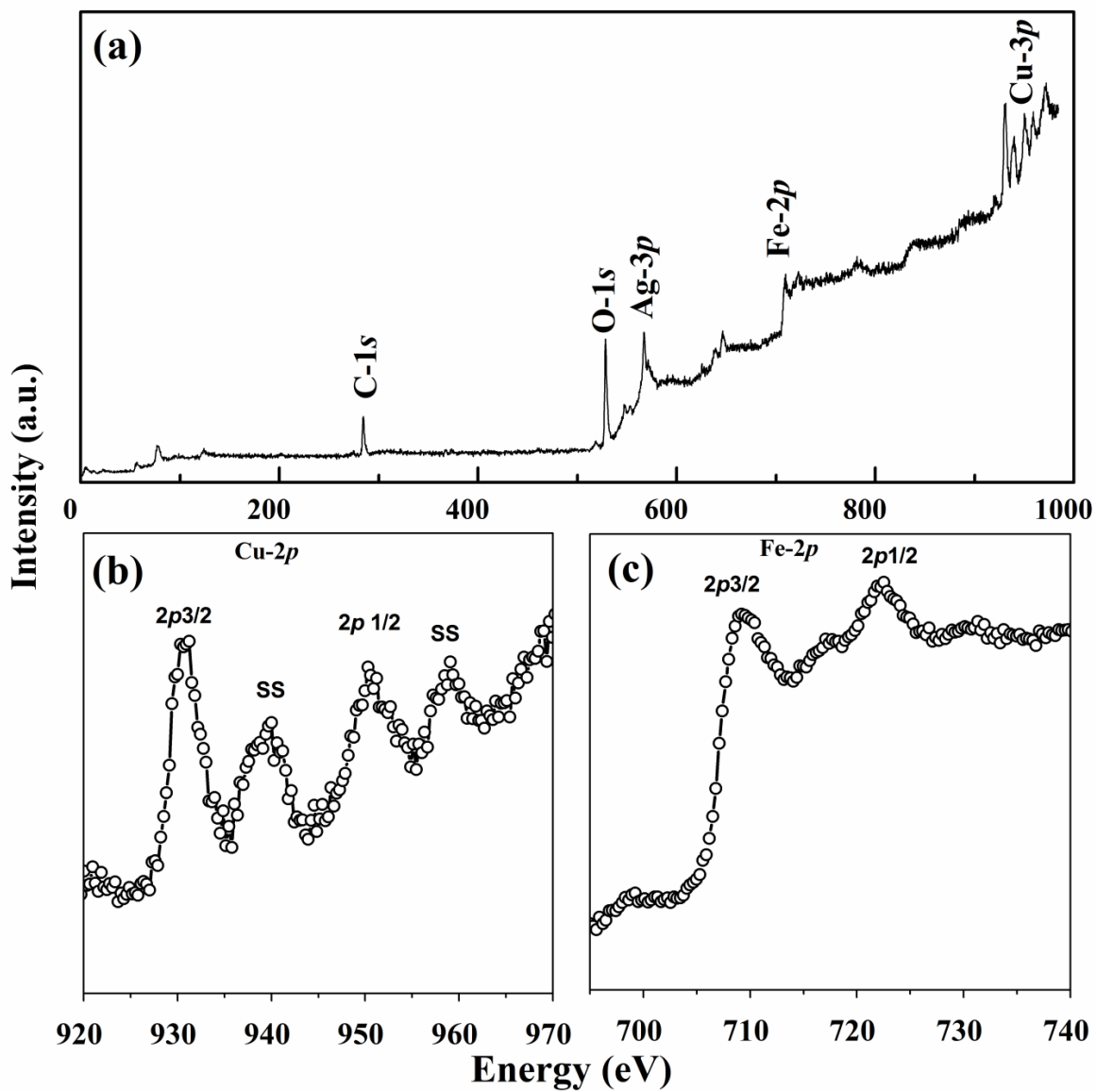

**Fig. 1.** (*a*) XPS Survey spectrum and high-resolution scans of (*b*) Cu2*p* and (*c*) Fe2*p* of Cu$_{0.875}$Fe$_{0.125}$O Nanoparticles.

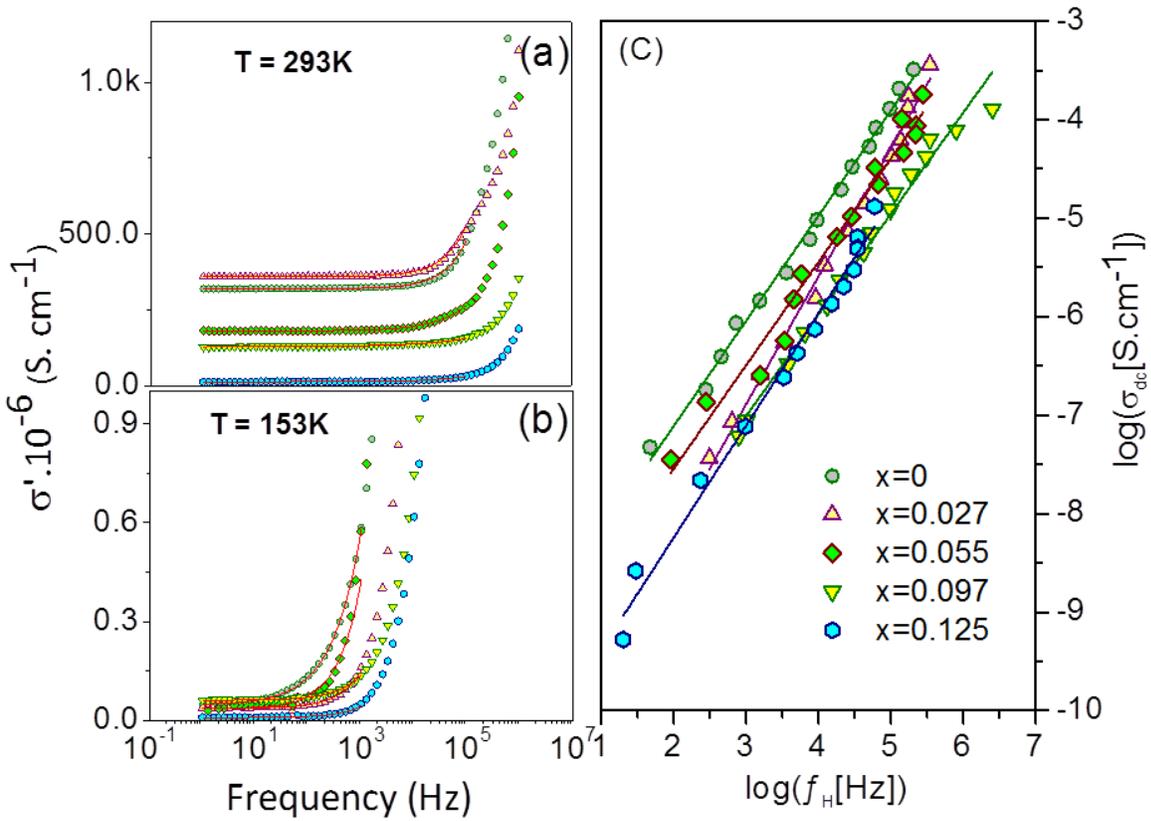

**Fig. 2.** Frequency spectra of the real conductivity (σ') at (a) 293 K, (b) 153 K, and (c) BNN plots for $Cu_{1-x}Fe_xO$ (0≤x≤0.125). The solid lines are obtained from a Jonscher's law fit of complex conductivities

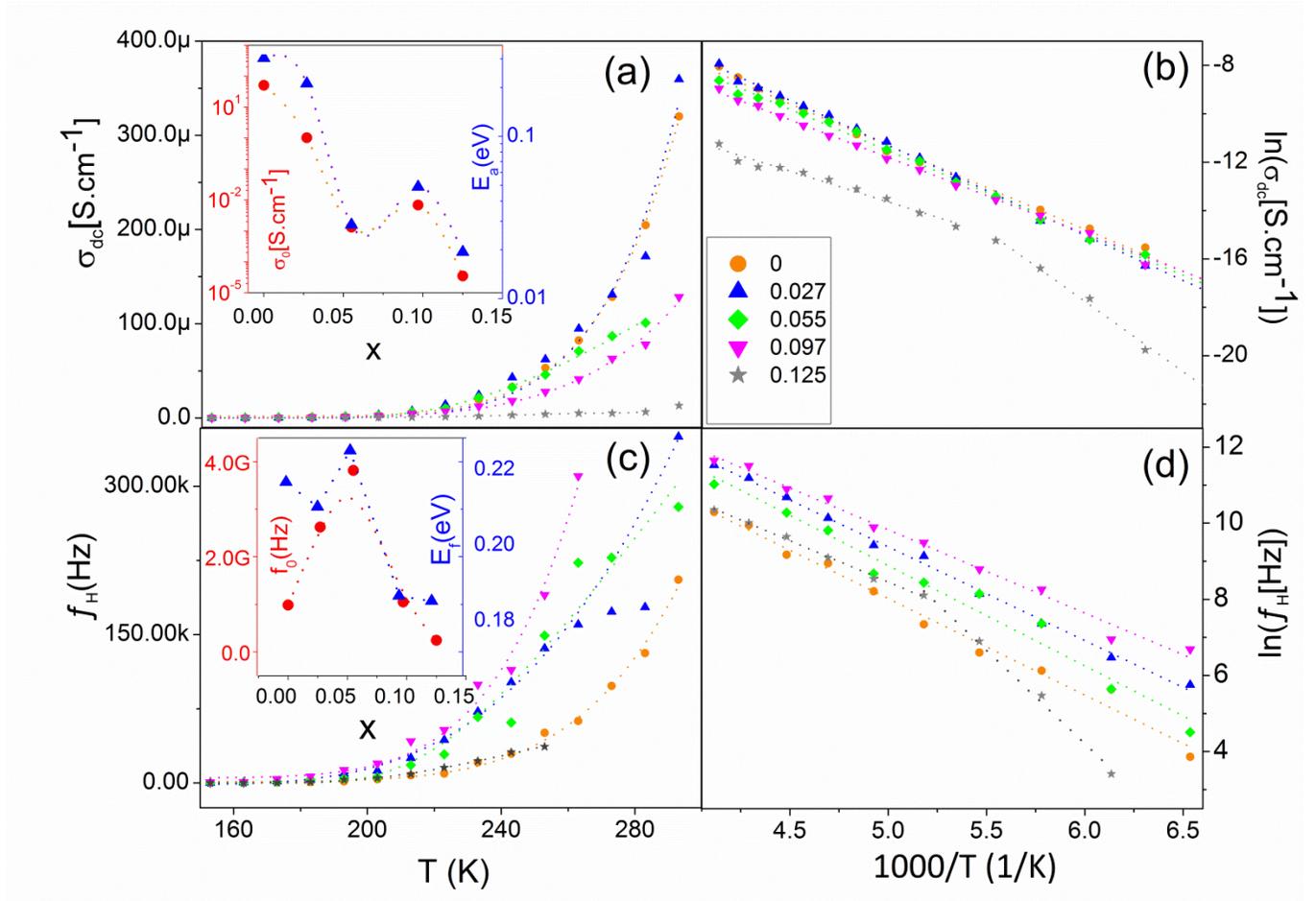

**Fig. 3.** Temperature dependence of (a) dc conductivity, $\sigma_{dc}$ vs T, (b) log($\sigma_{dc}$) vs 1000/T, (c) hopping frequency, $f_H$ vs T, and (d) log($f_H$) vs 1000/T for $Cu_{1-x}Fe_xO$ ($0 \leq x \leq 0.125$). The dotted lines represent fitted spectra.

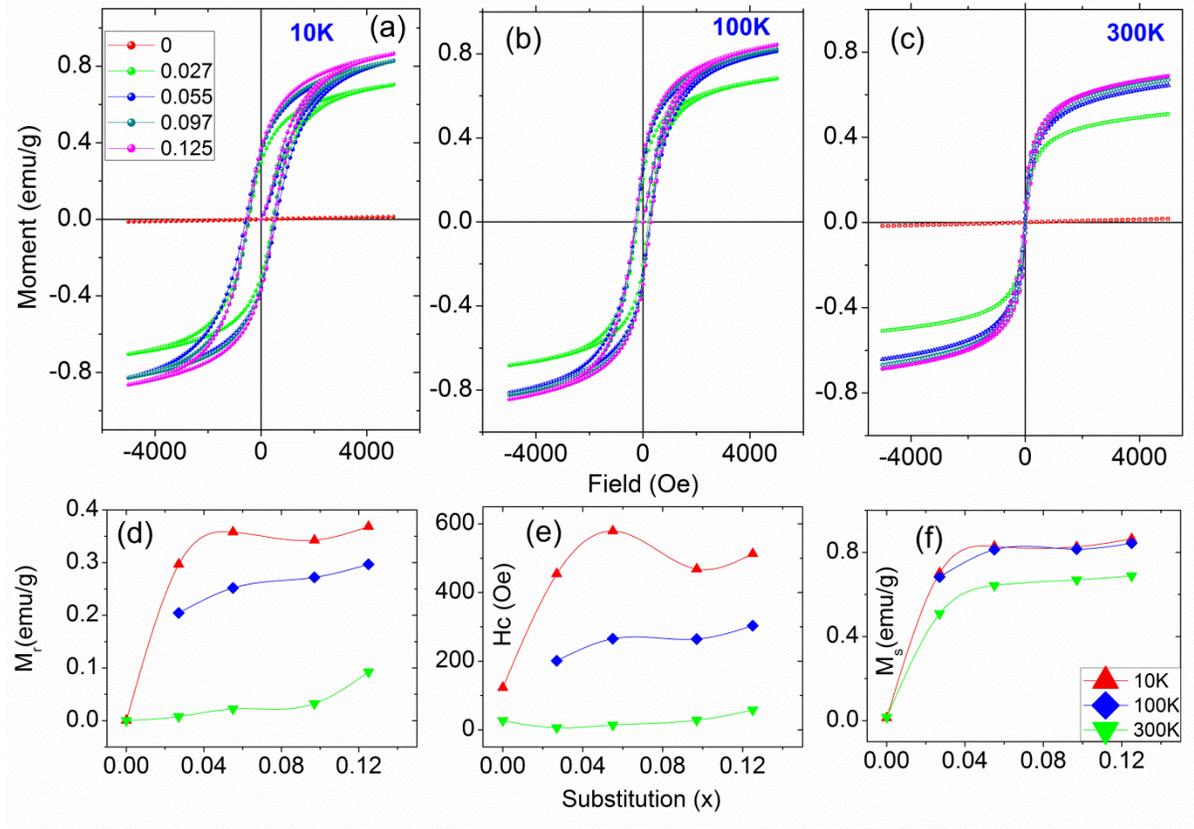

**Fig. 4.** Ferromagnetic hysteresis loops (M-H) at (a)10K, (b) 100K, (c)300K, (d)remnant magnetization ($M_r$), (e) coercivity ($H_c$), and (f) saturation magnetization ($M_s$) vs substitution for $Cu_{1-x}Fe_xO$ ($0 \leq x \leq 0.125$) samples.